\RequirePackage{fix-cm}
\documentclass[smallextended]{svjour3}       
\smartqed  
\usepackage{graphicx}

\usepackage{tikz}

\journalname{jsnm}
\begin{document}

\title{Cooper pair splitting efficiency in the hybrid three-terminal quantum dot
\thanks{This work is partially supported by the National Science Centre in Poland
via the project DEC-2014/13/B/ST3/04451.}
}


\author{Grzegorz Micha{\l}ek \and Tadeusz Doma\'{n}ski \and Karol I. Wysoki\'{n}ski}
\institute{G. Micha{\l}ek \at Institute of Molecular Physics, Polish Academy of Sciences,
60-179 Pozna\'{n}, Poland \\
T. Doma\'{n}ski, K. I. Wysoki\'{n}ski \at Institute of Physics, M. Curie-Sk{\l}odowska University, 20-031 Lublin, Poland \\ Tel.: +48-81 537 6236 \\ Fax: +48-81 537 6190 \\ \email{karol.wysokinski@poczta.umcs.lublin.pl}}




\date{Received: date \today / Accepted: date}

\maketitle

\begin{abstract}
Nanodevices consisting of a quantum dot tunnel coupled to one superconducting and two normal electrodes may serve as a source of entangled electrons. As a result of crossed Andreev reflection the Cooper pair of s-wave character may be split into two electrons and each of them goes into a distinct normal electrode, preserving entanglement. Efficiency of the process depends on the specific system and is tunable by electric means. Our calculations show that in the studied device this efficiency may attain values as large as $80 \%$.
\keywords{Cooper pair splitting \and hybrid structures \and Andreev reflection}
 \PACS{73.23.-b \and 73.63.Kv \and 74.45.+c}
\end{abstract}

\section{Introduction} \label{intro}

One of the important motivations to study hybrid many-terminal nanostructures with normal and superconducting leads is a perspective to obtain the non-locally entangled electrons~\cite{horodecki2009}.  Electrons of s-wave superconductors are known to exist in the spin singlet, maximally entangled states. When two electrons forming such Cooper pair are split (on interface between superconductor and other materials), they still preserve entanglement albeit being spatially separated. Splitting mechanism can be achieved~\cite{torres1999,recher2001} by the Andreev scattering processes. In many-terminal hybrid device with one superconducting electrode (Fig.~\ref{fig1}a) the crossed Andreev reflection (CAR) occurs when one of the electrons enters, say left (L) electrode while its partner goes to the right (R) one. The Cooper pair splitting (CPS) in such Y-shape junctions has been proposed theoretically~\cite{torres1999,recher2001,chtchelkatchev2002}
and realised experimentally by several groups~\cite{hofstetter2009,herrmann2010,wei2010,hofstetter2011,schindele2012,das2012,schindele2014}.

It has been suggested that the highest efficiency of CPS can be achieved in three-terminal hybrid devices using the double quantum dots with strong intradot Coulomb repulsion ~\cite{recher2001}. We have previously analysed the Y-shape junction with only a single quantum dot~\cite{michalek2013}, where interplay between the local and non-local processes is controlled by suitable choice of
the quantum dot level, the couplings and applied voltages \cite{michalek2015}.
The analysis of the thermoelectric properties~\cite{wysokinski2012,michalek2016} 
of this three terminal hybrid system has shown that Seebeck coefficient can be large, but the
symmetry of the model prevents a direct contribution of the CAR to both the local 
and non-local thermopower~\cite{michalek2016}.

Regions in the parameter space, where CAR processes are dominant under realistic experimental conditions~\cite{michalek2013,michalek2015,michalek2016} are expected to be also useful for the efficient CPS. It is the aim of this work to quantify such conjecture by calculating CPS efficiency as defined
in Eq. (\ref{CPS_definition}) below. The paper is organised as follows. In section \ref{sec:1} we introduce the model, discuss the approach and define the CPS efficiency. Our results are presented in section~\ref{sec:2} and in section~\ref{sec:3} we end with the brief summary.

\section{The system, its modelling and efficiency of CPS} \label{sec:1}

\begin{figure}
\includegraphics[width=0.49\linewidth]{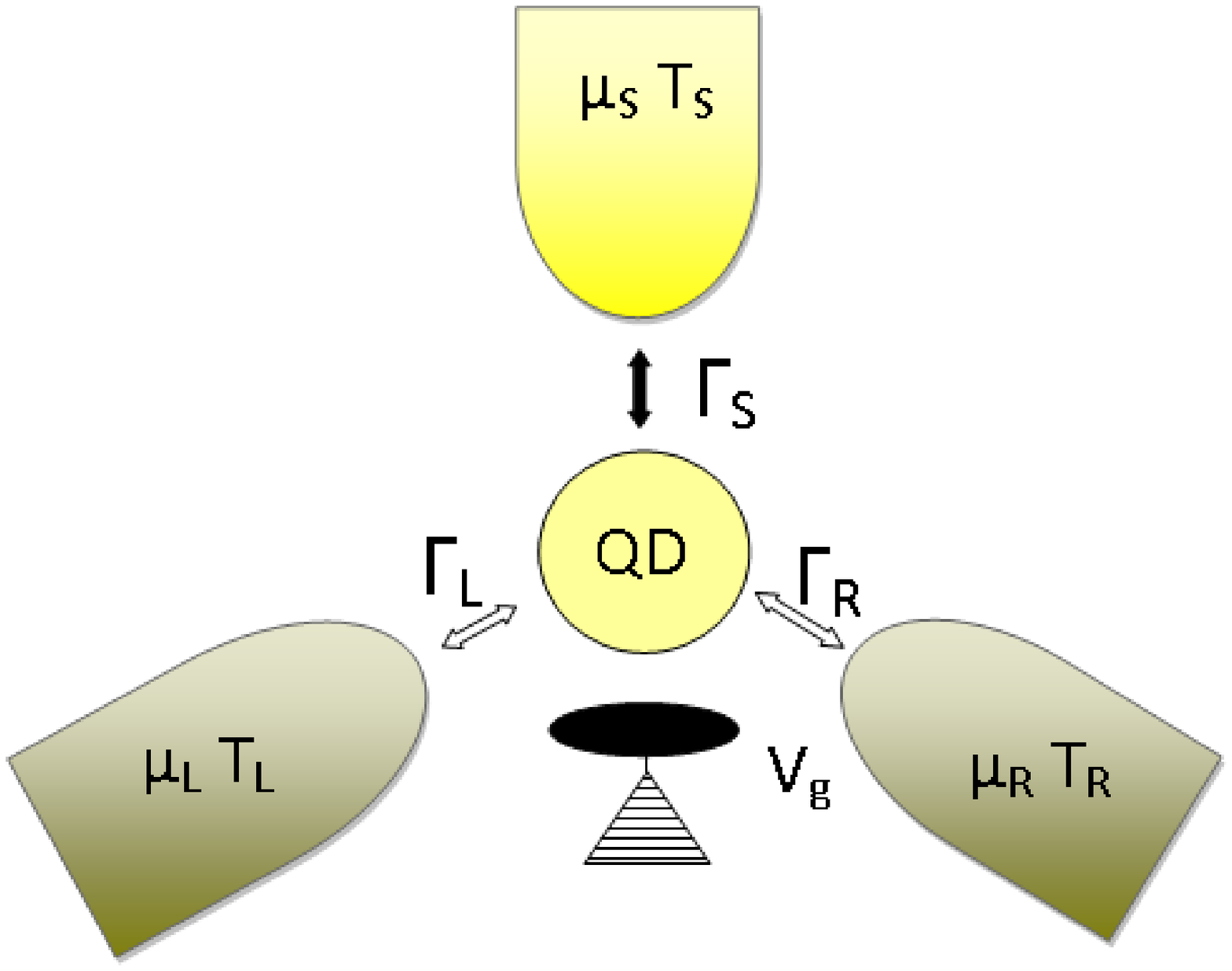}
\includegraphics[width=0.49\linewidth]{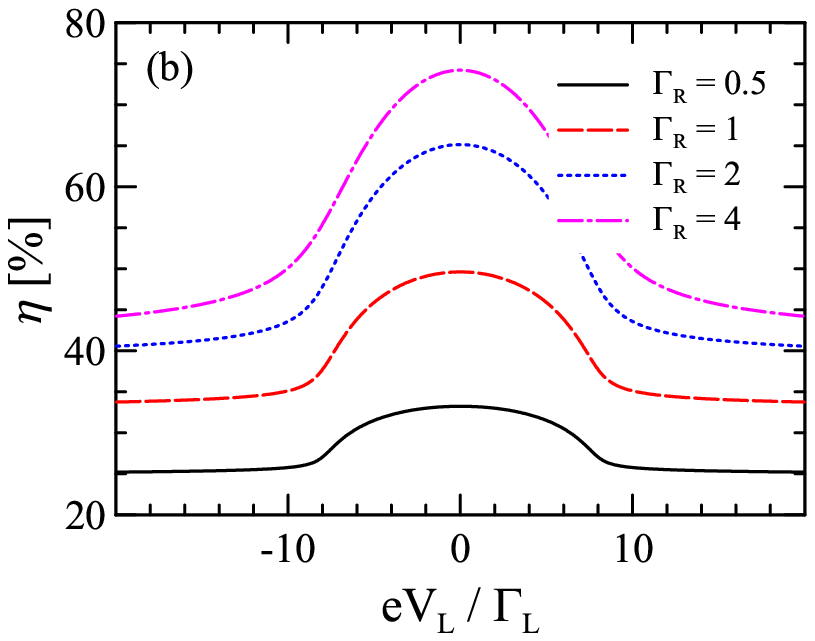}
\caption{(a) Sketch of the system. (b) Efficiency $\eta$ with respect to $V_L$ for $\Gamma_S / \Gamma_L = 16$, $\epsilon_0 = 0$, $U = 0$, $V_R = 0$, $T = 0$ and $\Gamma_R / \Gamma_L = \{ 0.5, 1, 2, 4 \}$. $\Gamma_L$ is treated as unity in our calculations.} \label{fig1}
\end{figure}

We consider the system, consisting of a quantum dot (QD) coupled to two normal (left - L and right - R) electrodes and another superconducting (S) lead. Such heterostructure (Fig.~\ref{fig1}a) can be modelled by the Hamiltonian
\begin{equation} \label{eq-Ham}
H = H_{QD} + \sum_{\alpha = L, R, S} H_\alpha + H_T \; ,
\end{equation}
where $H_{QD}$ describes the quantum dot, $H_\alpha$ refers to electrons of $\alpha$-th lead and $H_T$ is a hybridisation between the leads and the QD. Various parts of the Hamiltonian read $H_{QD} = \epsilon_0 \sum_\sigma d_\sigma^\dag d_\sigma + U n_\uparrow n_\downarrow$, where $\epsilon_0$ is the single-particle energy level, $d_\sigma^\dag$ ($d_\sigma$) denotes creation (annihilation) operator of the dot electron with spin $\sigma$, $n_\sigma \equiv d_\sigma^\dag d_\sigma$ is the number operator, and $U$ is the repulsive Coulomb interaction. The normal metal electrodes are treated within the mean field approximation $H_\alpha = \sum_{k, \sigma} \epsilon_{\alpha k} c_{\alpha k \sigma}^\dag c_{\alpha k \sigma}$, where $c_{\alpha k \sigma}^\dag$ ($c_{\alpha k \sigma}$) denotes creation (annihilation) of an electron with spin $\sigma$ and momentum $k$ in the electrode $\alpha = \{ L, R \}$.
The superconducting electrode is described in the BCS approximation by $H_S = \sum_{k, \sigma} \epsilon_{S k} c_{S k \sigma}^\dag c_{S k \sigma} + \sum_k \left( \Delta c_{S -k \uparrow}^\dag c_{S k \downarrow}^\dag + \Delta^* c_{S k \downarrow} c_{S -k \uparrow} \right)$, where we have assumed the isotropic energy gap $\Delta$. Coupling between the QD and the external leads is given by $H_T = \sum_{\alpha, k, \sigma} \left(V_{\alpha, k} c_{\alpha k \sigma}^\dag d_\sigma + V_{\alpha, k}^* d_\sigma^\dag c_{\alpha k \sigma} \right)$, where $V_{\alpha, k}$ describes the hopping of an electron between QD and the state $k$ in the $\alpha$ lead. Electron and hole transfer between the QD and the leads is described by an effective tunneling rate $\Gamma_\alpha$, which in the wide-band approximation takes the form $\Gamma_\alpha = 2 \pi \sum_k |V_{\alpha, k}|^2 \delta (E - \epsilon_{\alpha k})$ and is assumed to be energy independent. We shall consider finite voltage $V_L$ applied to the left electrode, whereas the superconducting and right electrodes are grounded.

In this work we are predominantly interested in the CPS, therefore we assume that the voltage $|V_{L}|$ is safely smaller than the superconducting gap. Focusing on the subgap transpor we determine the current $I_\alpha^{TOT}$ flowing from the normal $\alpha$ electrode~\cite{niu1999,sun1999}
\begin{equation} \label{tot-cur-L}
I_\alpha^{TOT} = I_\alpha^{ET} + I_\alpha^{AR} = I_\alpha^{ET} + I_\alpha^{DAR} + I_\alpha^{CAR} \; ,
\end{equation}
where
\begin{eqnarray} \label{ET-cur-L}
I_\alpha^{ET} &=& \frac{2e}{h} \Gamma_\alpha \Gamma_{\tilde{\alpha}} \int |G_{11}^r(E)|^2 \left[ f_\alpha(E) - f_{\tilde{\alpha}}(E) \right] dE \; , \\
\label{DAR-cur-L}
I_\alpha^{DAR} &=& \frac{2e}{h} \Gamma_\alpha^2 \int |G_{12}^r(E)|^2 \left[ f_\alpha(E) - \tilde{f}_\alpha(E) \right] dE \; , \\
\label{CAR-cur-L}
I_\alpha^{CAR} &=& \frac{2e}{h} \Gamma_\alpha \Gamma_{\tilde{\alpha}} \int |G_{12}^r(E)|^2 \left[ f_\alpha(E) - \tilde{f}_{\tilde{\alpha}}(E) \right] dE \; .
\end{eqnarray}
Here ${\tilde{\alpha}}$ denotes normal electrode different from $\alpha$ while $f_\alpha (E) = \{ \exp[( E - e V_\alpha ) / k_B T_\alpha] + 1 \}^{-1}$ and $\tilde{f}_\alpha (E) = 1 - f_\alpha(-E) = \{ \exp[( E + e V_\alpha ) / k_B T_\alpha] + 1 \}^{-1}$ are the Fermi-Dirac distribution functions in the electrode $\alpha = \{ L, R \}$ for electrons and holes, respectively. $I_\alpha^{ET}$ denotes the current contributed by the normal electron transfer (ET) and $I_\alpha^{AR}$ stands for the Andreev reflection (AR) processes due to the direct (DAR) and the crossed (CAR) scatterings. The current flowing  from the S-electrode is denoed $I_S^{TOT}$. The Kirchoff's law $I_L^{TOT} + I_R^{TOT} + I_S^{TOT} = 0$ is fulfilled.

In order to establish the optimal CPS efficiency we guide ourselves by the previous stidies~\cite{michalek2013,michalek2015}, looking for the model parameters where the CAR processes are dominant.
We use the following definition of CPS efficiency
\begin{equation}
\eta =
\frac{2 |I^{CAR}|}{2 |I^{ET}| + |I_L^{DAR}| + |I_R^{DAR}| + 2 |I^{CAR}|} \; ,
\label{CPS_definition}
\end{equation}
where $I^{CAR} \equiv I_L^{CAR} = I_R^{CAR}$ and $I^{ET} \equiv I_L^{ET} = - I_R^{ET}$. The expression (\ref{CPS_definition}) generalises $\eta = 2 |I^{CAR}|  / \left( 2 |I^{CAR}| + |I_L^{DAR}| + |I_R^{DAR}| \right)$ previously introduced under isothermal conditions for $V_L = V_R \neq V_S$ \cite{herrmann2010} and is consistent with the definition $\eta =  |I^{CAR}| / \left( |I^{CAR}| + |I^{ET}| \right)$ used in absence of any electrostatic voltage for $T_L \neq T_R$ \cite{cao2015}. In what follows we investigate the CPS efficiency (\ref{CPS_definition}) for producing the entangled electrons as a consequence of competition between the CAR and other transport channels.

\section{Results} \label{sec:2}

\begin{figure}
\includegraphics[width=\linewidth,clip]{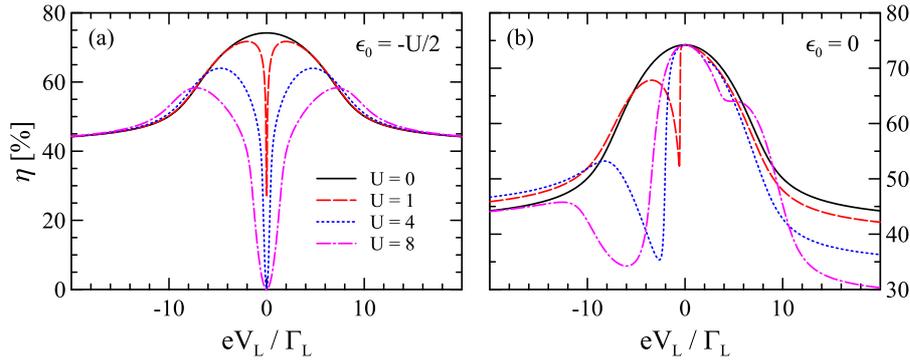}
\caption{Dependence of the CPS efficiency $\eta$ on the voltage $V_L$ for $\Gamma_S / \Gamma_L = 16$, $\Gamma_R / \Gamma_L = 4$, $V_R = 0$, $T = 0$, $U / \Gamma_L = \{ 0, 1, 4, 8 \}$ and (a) $\epsilon_0 = -U / 2$, (b) $\epsilon_0 = 0$.}
\label{fig2}
\end{figure}

\begin{figure}
\includegraphics[width=\linewidth,clip]{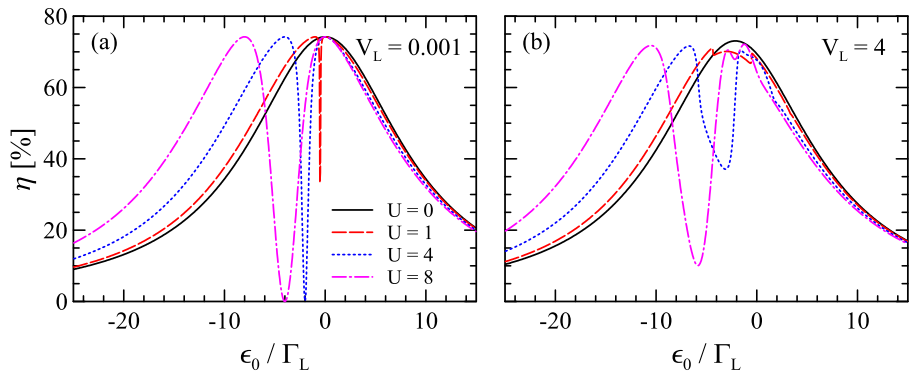}
\caption{Efficiency $\eta$ as a function of the quantum dot level $\epsilon_0$ for (a) small $e V_L / h \Gamma_L = 0.001$ and (b) large $e V_L / h \Gamma_L = 4$. The other parameters are the same as in Fig.~\ref{fig2}.}
\label{fig3}
\end{figure}

Figs~\ref{fig1}-\ref{fig3} show the results obtained numerically for voltages $V_R = V_S = 0$ and temperature $T = 0$. We have imposed the strong coupling condition to superconducting electrode  
assuming $\Gamma_S \gg \Gamma_{L, R}$ in order to promote the Andreev processes. For the sake of completeness we start the analysis with a simple noninteracting case $U = 0$. Fig.~\ref{fig1}b displays the CPS efficiency $\eta$ as a function of the voltage $V_L$ applied to the normal L electrode for QD level position $\epsilon_0 = 0$ and for various couplings to the normal R electrode. Magnitude of $\eta$ depends on $\Gamma_R$ and its flat maximum occurs for $V_L \rightarrow 0$, where the ET processes are suppressed because of the proximity induced on-dot gap. The CPS efficiency decreases when voltages exceeds the Andreev bound states (at energies $\pm \sqrt{\epsilon_0^2 + \Gamma_S / 4}$) which are formed in the QD due to the proximity effect~\cite{michalek2013}. The  increase of $\eta$ with  $\Gamma_R$ is caused by the CAR processes, dominant over the DAR currents (for $\Gamma_R > 2 \Gamma_L$). Let us note, however, that the ET current is also sensitive to $\Gamma_R$. This can be seen for large $\Gamma_R$  (not shown) when the CPS efficiency eventually decreases. Let us emphasize that for other electrical bias configurations (\textit{e.g.} $V_R = V_L$, $V_R = - V_L$) the CPS efficiency is reduced. For example, $\eta \rightarrow 0$ for $V_R = -V_L$ and $T \rightarrow 0$ due to the strong suppression of the CAR processes.

Fig.~\ref{fig2} presents the CPS efficiency (\ref{CPS_definition}) obtained for the correlated quantum dot, where the Coulomb interactions are treated in the Hubbard I approximation (\textit{e.g.}~\cite{haug-jauho1996,michalek2013}). For finite $U$, in contrast to the noninteracting case,  four Andreev bound states~\cite{michalek2013} appear in the density of states of the QD. For the electron-hole symmetric case $\epsilon_0 = -U/2$ the CPS efficiency $\eta$ is a symmetrica function of the applied bias voltage $V_L$. The CPS efficiency diminishes with an increasing Coulomb repulsion. For small voltages, when system is in the Coulomb blockade regime (\textit{i.e.} in the region between the inner Andreev bound states) there appears a dip, where $\eta$ is reduced. It is in contrast with the noninteracting case where the maximum of $\eta$ appeared between Andreev bound states. This behaviour is manifestation of the Coulomb blockade effect, which suppresses the CAR processes more efficiently then the ET processes~\cite{michalek2015}. As the separation of the Andreeev bound states grows with an increase of the electron interactions $U$ the corresponding distance between maxima of the CPS efficiency also increases. When the electron-hole symmetry is broken (\textit{e.g.} for $\epsilon_0 = 0$) we can notice that the CPS efficiency characteristics are asymmetric with respect to the $V_L = 0$. This behavior is caused by the ET contribution, which amplitude is sensitive on the position of the $\epsilon_0$. We can also observe a dip, which is less pronounced and slightly shifted to larger voltages $|V_L|$ while the magnitude and position of the maximum remains unchanged.

Variation of the CPS efficiency $\eta$ with respect to the quantum dot level $\epsilon_0$ is shown in Fig.~\ref{fig3}. For small bias $V_L / \Gamma_L = 0.001$ the optimal values of $\eta$ appear symmetrically with respect to the particle-hole symmetry point  $\epsilon_0 \simeq -U/2$ where $\eta$
vanishes. The height of both peaks remain unchanged by the Coulomb potential, in analogy to the recent observation of CPS in the Josephson junction\cite{deacon2015}. For larger bias $V_L / \Gamma_L = 4$ the situation is a bit more complicated as  the pronounced dip between the peaks appears for sufficiently large electron-electron interactions. Position of the maxima are slightly shifted relative to the small bias case. The optimal CPS efficiency can by precisely tuned by the gate voltage which shifts the energy level $\varepsilon_0$ and by the bias voltage $V_L$. 
This complex behaviour of $\eta$ is driven by a competition between the ET, DAR and CAR  processes in accord with earlier discussion~\cite{michalek2013}.

\section{Summary} \label{sec:3}

We have extended our previous studies~\cite{michalek2013,michalek2015,michalek2016} by analysing the efficiency of the Cooper pair splitting in a hybrid system, consisting of the quantum dot coupled to one superconducting and two metallic electrodes. We have found that the CPS efficiency is tunable (by changing the system parameters) and its optimal value can approach nearly $80 \%$. The repulsive interactions have the destructive influence on such efficiency. In the Coulomb blockade regime the CPS efficiency is suppressed  due to the reduced probability to find a pair of electrons on the quantum dot.

In distinction to Ref.\ \cite{cao2015} (corresponding to the Y-shape heterostructure with two quantum dots) we have found that our system does not allow for any non-zero CPS efficiency originating solely from the temperature difference (i.e.\ in absence of any external voltage). This effect is caused by the particle-hole symmetry of the crossed Andreev reflections,
as has been recently emphasized in a context on the thermoelectric properties \cite{michalek2016}.


\begin{thebibliography}{}
%

\bibitem{horodecki2009}
R. Horodecki, P. Horodecki, M. Horodecki, and K. Horodecki, Rev. Mod. Phys. \textbf{81}, 865 (2009).

\bibitem{torres1999} J. Torr\'{e}s and T. Martin, Eur. Phys. J. B \textbf{12}, 319 (1999).

\bibitem{recher2001} P. Recher, E. V. Sukhorukov, and D. Loss, Phys. Rev. B \textbf{63}, 165314 (2001).

\bibitem{chtchelkatchev2002} N. M. Chtchelkatchev, G. Blatter, G. B. Lesovik, and T. Martin, Phys. Rev. B \textbf{66}, 161320(R) (2002).

\bibitem{hofstetter2009} L. Hofstetter, S. Csonka, J. Nyg{\aa}rd, and C. Sch\"{o}nenberger, Nature \textbf{461}, 960 (2009).

\bibitem{herrmann2010} L. G. Herrmann, F. Portier, P. Roche, A. L. Yeyati, T. Kontos, and C. Strunk, Phys. Rev. Lett. \textbf{104}, 026801 (2010).

\bibitem{wei2010} J. Wei and V. Chandrasekhar, Nat. Phys. \textbf{6}, 494 (2010).

\bibitem{hofstetter2011} L. Hofstetter, S. Csonka, A. Baumgartner, G. F\"{u}l\"{o}p, S. d'Hollosy, J. Nyg{\aa}rd, and C. Sch\"{o}nenberger, Phys. Rev. Lett. \textbf{107}, 136801 (2011).

\bibitem{schindele2012} J. Schindele, A. Baumgartner, and C. Sch\"{o}nenberger, Phys. Rev. Lett. \textbf{109}, 157002 (2012).

\bibitem{das2012} A. Das, Y. Ronen, M. Heiblum, D. Mahalu, A. V. Kretinin, and H. Shtrikman, Nat. Commun. \textbf{3}, 1165 (2012).

\bibitem{schindele2014} J. Schindele, A. Baumgartner, R. Maurand, M. Weiss, and C. Sch\"{o}nenberger, Phys. Rev. B \textbf{89}, 045422 (2014).

\bibitem{michalek2013}
G. Micha{\l}ek, B. R. Bu{\l}ka, T. Doma\'{n}ski, and K. I. Wysoki\'{n}ski, Phys. Rev. B \textbf{88}, 155425 (2013).

\bibitem{michalek2015}
G. Micha{\l}ek, T. Doma\'{n}ski, B. R. Bu{\l}ka, and K. I. Wysoki\'{n}ski, Sci. Rep. \textbf{5},  14572 (2015).

\bibitem{wysokinski2012}
K. I. Wysoki\'{n}ski, J. Phys.: Condens. Matter \textbf{24}, 335303 (2012).


\bibitem{michalek2016}
G. Micha{\l}ek, M. Urbaniak, B. R. Bu{\l}ka, T. Doma\'{n}ski, and K. I. Wysoki\'{n}ski
Phys. Rev. B \textbf{93}, 235440 (2016).

\bibitem{niu1999} C. Niu, D. L. Lin, and T.-H. Lin, J. Phys.: Condens. Matt. \textbf{11}, 1511 (1999).

\bibitem{sun1999} Q.-F. Sun, J. Wang, and T.-H. Lin, Phys. Rev. B \textbf{59}, 3831 (1999); Phys. Rev. B \textbf{62}, 648 (2000); Y. Zhu, T.-H. Lin, and Q.-F. Sun, Phys. Rev. B \textbf{69}, 121302 (2004).

\bibitem{cao2015}
Z. Cao, T. F. Fang, L. Li, and H. G. Luo, Appl. Phys. Lett. \textbf{107}, 212601 (2015).

\bibitem{haug-jauho1996}
H. Haug and A.-P. Jauho, \textit{Quantum Kinetics in Transport and Optics of Semiconductors, Second, Substantailly Revised Edition}, Springer Verlag, Berlin (2008).

\bibitem{deacon2015}
R. S. Deacon, A. Oiwa, J. Sailer, S. Baba, Y. Kanai, K. Shibata, K. Hirakawa,
and S. Tarucha, Nat. Commun. \textbf{6}, 7446 (2015).

\end{thebibliography}


\end{document}